\documentclass{article} % use larger type; default would be 10pt

\usepackage[utf8]{inputenc} % set input encoding (not needed with XeLaTeX)

\usepackage[margin=1in,a4paper]{geometry} % to change the page dimensions

\usepackage{amsmath, amsthm, amssymb}

\usepackage{comment}
\usepackage{microtype}
\usepackage{todonotes}
\usepackage{xspace}
\usepackage{authblk}
\usepackage[hidelinks]{hyperref}
\usepackage[linesnumbered,ruled,noend]{algorithm2e}
\usepackage{enumitem}
\usepackage{tikz}
\usepackage{tikz-cd}
\usepackage{standalone}

\usepackage[capitalize]{cleveref}

\newcommand{\inborder}{input border\xspace}
\newcommand{\inborders}{input borders\xspace}
\newcommand{\outborder}{output border\xspace}
\newcommand{\outborders}{output borders\xspace}
\newcommand{\outputtop}{\ensuremath{\operatorname{OUT}_{\operatorname{TOP}}}\xspace}
\newcommand{\outputleft}{\ensuremath{\operatorname{OUT}_{\operatorname{LEFT}}}\xspace}
\newcommand{\inputtop}{\ensuremath{\operatorname{TOP}}\xspace}
\newcommand{\inputleft}{\ensuremath{\operatorname{LEFT}}\xspace}

\DeclareMathOperator{\dom}{dom}

\newcommand{\Oh}{\mathcal{O}}
\newcommand{\SWM}{\textsf{SWM}\xspace}

\newtheorem{thm}{Theorem}[section]
\newtheorem{theorem}[thm]{Theorem}

\newtheorem{lemma}[thm]{Lemma}

\newtheorem{observation}[thm]{Observation}

\newtheorem{definition}[thm]{Definition}

\crefname{corollary}{Corollary}{Corollaries}
\crefname{fact}{Fact}{Facts}

\title{RLE edit distance in near optimal time}
\author[1]{Rapha\"el Clifford}
\author[2]{Paweł Gawrychowski}
\author[,3,4]{Tomasz Kociumaka\thanks{Supported by ISF grants no. 824/17 and 1278/16 and by an ERC grant MPM under the EU's Horizon 2020 Research and Innovation Programme (grant no. 683064).}}
\author[5,6]{Daniel P. Martin}
\author[2]{Przemysław Uznański}

\affil[1]{Department of Computer Science, University of Bristol, UK} 
\affil[ ]{\texttt{Raphael.Clifford@bristol.ac.uk}}
\affil[2]{Institute of Computer Science, University of Wrocław, Poland}
\affil[ ]{\texttt{\{gawry,puznanski\}@cs.uni.wroc.pl}}
\affil[3]{Department of Computer Science, Bar-Ilan University, Israel}
\affil[4]{Institute of Informatics, University of Warsaw, Poland}
\affil[ ]{\texttt{kociumaka@mimuw.edu.pl}}
\affil[5]{School of Mathematics, University of Bristol, UK}
\affil[6]{Heilbronn Institute for Mathematical Research, Bristol, UK}
\affil[ ]{\texttt{Dan.Martin@bristol.ac.uk}}

\date{\vspace{-1.5cm}}

\begin{document}
\maketitle
\begin{abstract}

 We show that the edit distance between two run-length encoded strings of compressed lengths $m$ and $n$ respectively, can be computed in $\mathcal{O}(mn\log(mn))$ time.    This improves the previous record by a factor of $\mathcal{O}(n/\log(mn))$. The running time of our algorithm is within subpolynomial factors of being optimal, subject to the standard SETH-hardness assumption. This effectively closes a line of algorithmic research first started in 1993.
\end{abstract}

% !TEX root = main.tex
\section{Introduction}
 
The edit distance is one of the most common distance measures between strings.  For two strings of length $M$ and $N$ respectively, the edit distance counts the minimum number of single character insertions, deletions and substitutions needed to transform one string into the other. The first record of an $\Oh(MN)$ algorithm to compute the edit distance is from 1968~\cite{Vintsyuk:1968} although it was rediscovered independently a number of times subsequently.   Masek and Paterson improved the running time to $\Oh(MN/\log{M})$ in 1980 and this is the fastest known algorithm to date~\cite{Masek:1980}.  Much more recently it has been shown no $\Oh(MN^{1-\epsilon})$ time edit distance algorithm can exist, subject to the strong exponential time hypothesis (SETH)~\cite{BI:SETH:2015,BK:2015}. As a result, it is likely that little further progress can be made in terms of improving its worst case complexity.

In this paper we focus on the problem of computing the edit distance between two compressed strings. The run-length encoding (RLE) of a string compresses  consecutive identical symbols into a run, denoted $\sigma^i$ if the symbol $\sigma$ is repeated $i$ times.  For example $\texttt{aaabbbbaaa}$ would be compressed to $\texttt{a}^3\texttt{b}^4\texttt{a}^3$.  This form of compression is commonly used for image compression but also has wider applications including, for example, in image processing~\cite{HFD:1990,XKFR:2004} and succinct data structures~\cite{MN:2005}.

In 1993 Bunke and Csirik proposed the first algorithm for computing the edit distance between RLE strings. For two strings of RLE-compressed lengths $m$ and $n$ respectively, their algorithm runs in $\Oh(mn)$ time in the special case where all the runs are of the same length~\cite{BC:1993}.  However the running time falls back to the naive complexity of $\Oh(MN)$ time in the worst case where $M$ and $N$ are the uncompressed lengths of the two strings.  This worst case complexity was subsequently improved to $\Oh(Nm + Mn)$~\cite{BC:1995,ALM:2002} and then $\Oh(\min\{Nm, Mn\})$ time in 2007~\cite{LGYL:2007}. Finally in 2013 the fastest solution prior to this current work was given running in $\Oh(mn^2)$ time, where $n\geq m$~\cite{CC-RLEedit:2013}. This was the also the first algorithm for the RLE edit distance problem whose running time did not depend on the uncompressed lengths of the input strings.

For uncompressed strings, the longest common subsequence (LCS) problem has long been considered a close relative of the edit distance problem. This is partly due to the similarity of their dynamic programming solutions and partly because LCS is a special case of edit distance when general costs are allowed for the different mismatch and substitution operations.  Moreover, the two problems have the same quadratic time upper bounds and SETH-hardness lower bounds~\cite{BK:2015}. Somewhat surprisingly, however, the history of algorithms for LCS and edit distance have \emph{not} mirrored each other when the problems are considered on RLE strings. In particular, an $\Oh(mn\log(mn))$ time algorithm for computing the LCS on RLE strings was given in 1999~\cite{ALS:1999} which is considerably faster than has been possible up to this point for the edit distance problem. Some work has also been carried out since that date to improve the log factor in the running time complexity for the LCS problem~\cite{AYTH:2008,Sakai:2012}.

In this paper we speed up the running time for the edit distance problem on RLE strings by a factor of $\mathcal{O}(n/\log(mn))$, matching the fastest LCS algorithm to within a logarithmic factor and making it within subpolynomial factors of being optimal, assuming SETH holds. As a result, our new algorithm shows that the LCS and edit distance problems are indeed of essentially the same complexity even when the input strings are run-length encoded.  This effectively closes a line of algorithmic research first started in 1993.
\begin{theorem}
	Given two RLE strings of compressed length $n$ and $m$ respectively, there exists an algorithm to compute their edit distance which runs in $\mathcal{O}(mn\log(mn))$ time.
\end{theorem}\label{thm:one}

% !TEX root = main.tex
\section{Previous Work and Preliminaries}\label{sec:previous}
The classic dynamic programming solution for computing the edit distance between uncompressed strings $X$ and $Y$ of uncompressed lengths $M$ and $N$ respectively, computes the  distance between all prefixes $X[1,\dots,i]$ and $Y[1,\dots,j]$.   The key recurrence which enables us to do this efficiently is given by: 
\begin{align*}\textsf{ED}(i,j) = \min(&\textsf{ED}(i-1, j-1) + \delta(X_{i} \ne Y_{j}), \textsf{ED}(i-1,j) + 1, \textsf{ED}(i,j-1) + 1).
\end{align*}
From this the classic $O(MN)$ time solution follows directly.   

\begin{figure}[b!]
		\centering
\begin{tikzpicture}
\draw (0, 0) rectangle (10, 5);
\draw[fill=gray!50] rectangle (2.5, 2);
\draw[fill=gray!50] (0, 4) rectangle (2.5, 5);
\draw[fill=gray!50] (4, 0) rectangle (8, 2);
\draw[fill=gray!50] (4, 4) rectangle (8, 5);
\draw[fill=gray!50] (2.5, 2) rectangle (4, 4);
\draw[fill=gray!50] (8, 2) rectangle (10, 4);
\draw[densely dotted, line width=2.5pt] (4, 2.05) -- (8, 2.05);
\draw[densely dotted, line width=2.5pt] (7.95, 2) -- (7.95, 4);
\draw[line width=2.5pt] (3.95, 2) -- (3.95, 4);
\draw[line width=2.5pt] (3.90, 4.05) -- (8, 4.05);
\node at (-0.3,1) {\texttt{a}};
\node at (-0.3,3) {\texttt{b}};
\node at (-0.3,4.5) {\texttt{a}};
\node at (1.25, 5.3) {\texttt{a}};
\node at (3.25, 5.3) {\texttt{b}};
\node at (6, 5.3) {\texttt{a}};
\node at (9, 5.3) {\texttt{b}};
\end{tikzpicture}
	\caption{Match blocks are in grey. The thick line is for an \inborder and the dotted line an \outborder. The \inborder is contained entirely inside the neighbouring match blocks.}\label{fig:borders}
\end{figure}
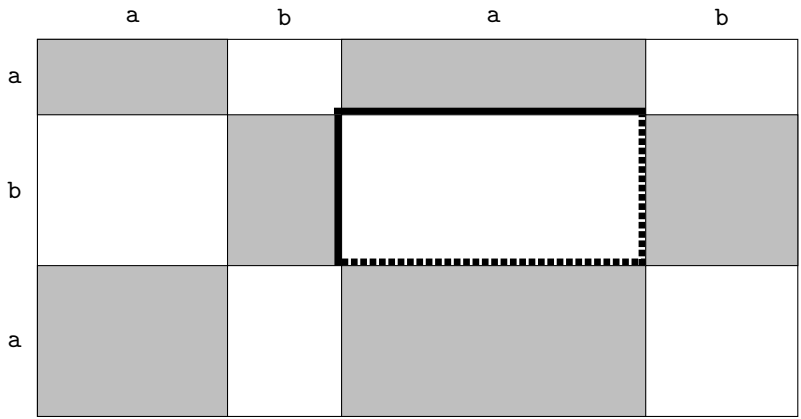

The previous approaches for the edit distance problem on RLE strings take this recurrence and the implied dynamic programming table as their starting point.  The basic idea was introduced by Bunke and Csirk~\cite{BC:1993} whose algorithm works by dividing the dynamic programming table
into ``blocks'', where each block is defined by a run in the original strings.

For each block the central task is to compute its bottom row and rightmost column given the bottom row of the block above and the rightmost column of the block to the left. For simplicity of terminology, we will refer to the rightmost column of the block to the left and the bottom row of the block above collectively as the \emph{\inborder} of a block and the bottom row and rightmost column of a block as its \emph{\outborder}. Figure~\ref{fig:borders} illustrates an example.

In~\cite{BC:1995,ALM:2002} it was shown that the work needed to derive the values of all the \outborders of blocks is at most linear in their length. When computing the edit distance between strings $X$ and $Y$, the length of each row in the dynamic programming table is the uncompressed length of string $Y$ and the length of each column is the uncompressed length of $X$. If there are $m$ runs in string $X$ and $n$ runs in $Y$ then the total time complexity for computing the edit distance using their approach is therefore $O(Nm + Mn)$.

%\todo{
%BC shows that you can compute the bottom and right quickly.
%CC shows that you only have to compute the turning points.
%Define turning point then quote lemma 11 from CC}

%Thus these can be computed efficiently. Therefore given two
%strings $X\in\Sigma^M,Y\in\Sigma^N$ with compressed
%$n$ and $m$ respectively, the algorithm has running time
%$O(Nm + Mn)$.

The work closest to ours is the $O(mn^2)$ algorithm of Chen and Chao~\cite{CC-RLEedit:2013}. They observe that the borders of the blocks in the dynamic programming table are piecewise linear with gradient $\pm1$ or $0$.
The borders can be therefore concisely represented by specifying their starting values as well as the positions and types of the points
of changing gradient called the \emph{turning points}.
They prove that for a given block the number of turning points in an \outborder is at most a constant greater than in its \inborder.
Consequently, a simple calculation shows that the total number of turning points is $\mathcal{O}(mn^2)$ (for $m\leq n$).
Chen and Chao arrive at their final complexity by designing a procedure that computes the representation of the \outborder of a block,
given the representation of its \inborder, in time proportional to the number of turning points.
We  now summarise their approach using our own notation.

There are two distinct types of blocks in the dynamic programming table. A match block corresponds to a rectangle where the corresponding symbols in the two strings match.  A mismatch block corresponds to a rectangle where the corresponding symbols mismatch.  Figure~\ref{fig:borders} shows both match and mismatch blocks. Borrowing notation from~\cite{CC-RLEedit:2013} we say that element $(i,j)$ of the dynamic programming table $\textsf{ED}$ is in \emph{diagonal} $j - i$.  Let $(i_d ,j_d)$ be the intersection of the \inborder with diagonal $j - i$.

\begin{lemma}[{\cite[Lemma 1]{CC-RLEedit:2013}}]\label{match-block}
	For a match block, $\textsf{ED}(i, j) = \textsf{ED}(i_d, j_d)$.
\end{lemma}

\cref{match-block} indicates that for a match block we can simply copy the values from the corresponding position in the \inborder to derive the values of the \outborder. The main challenge is therefore how to handle mismatch blocks.

For mismatch blocks Chen and Chao's algorithm, in a similar manner to previous RLE edit distance algorithms, splits the problem into two
parts corresponding to shortest paths that pass through the leftmost column or the top row~\cite{BC:1993,BC:1995,ALM:2002,CC-RLEedit:2013}.
Consider a mismatch block of height $h$ and width $w$ corresponding to runs $\texttt{a}^{h-1}$ and $\texttt{b}^{w-1}$
such that $h\leq w$ (the other case can be processed similarly by swapping the left and the top part of the \inborder).
$\inputleft[1,h]$ and $\inputtop[1,w]$ denotes the values of the left and the top
part of the input border, numbered in a bottom-to-top and a left-to-right direction, respectively.
For an array $S[1,n]$ and a parameter $h\in\mathbb{Z^+}$, let $S^{(h)}[i]=\min\{S[j]\mid i-h+1\leq j\leq i, 1\leq j\leq n\}$.
Chen and Chao separately compute all the \outborder values that are derived from a value in the left part of the \inborder,
denoted $\outputleft[1,w+h-1]$, and similarly compute all the \outborder values that are derived from a value
 in the top part of the input border, denoted $\outputtop[1,w+h-1]$, as follows.
 
 \begin{eqnarray}
 \outputleft & = & \begin{cases} \label{eqn:left}
 	\inputleft^{(h)}[i]+i-1 & \text{for } i\in [1,h]; \\
 	\inputleft^{(h)}[h]+i-1 & \text{for } i\in [h,w]; \\
 	\inputleft^{(h)}[i-w+h]+w-1 & \text{for } i\in [w,w+h-1];
 \end{cases} \\ 
 \outputtop & = & \begin{cases} \label{eqn:top}
 	\inputtop^{(h)}[i]+h-1 &\quad\text{for } i\in [1,w]; \\
 	\inputtop^{(h)}[i]+w+h-1-i &\quad\text{for } i\in [w,w+h-1].
 \end{cases}
 \end{eqnarray}

We start with reformulating the algorithmic framework of Chen and Chao using the following notation.

\begin{definition}\label{definitions}
  Let $S[1,n]$ be a 1-indexed array of length $n$.
  \begin{itemize}[itemsep=0ex, parsep=1pt, topsep=1pt]
  \item For a parameter $h\in\mathbb{Z^+}$, $\textsf{SWM}(S,h)$ returns the array $S^{(h)}$ of length $n+h-1$.
  \item $\textsf{split}(S, m)$ returns the two subarrays $S[1,m]$, and $S[m+1, n]$. 
  \item $S \pm\overrightarrow{1}$ returns $S$ with the gradient decreased/increased by one, or formally $S'[i]=S[i]\pm i$.
  \item For an integer constant $c$, $S + c$ returns $S$ with every value increased by $c$.
  \item $\textsf{initialise}(\ell)$ returns an array of length $\ell$ initially filled with zeroes.
  \item $\textsf{join}(S_{1},S_{2})$ simply concatenates two arrays.
  \end{itemize}
\end{definition}

\begin{algorithm}[b]

	$S\; \leftarrow \textsf{SWM}_h(\inputleft[1,h])$\;
	$S^\ell, S^r \leftarrow \textsf{split}(S,h)$\;
	$S_1 \leftarrow S^\ell + \overrightarrow{1} -1$\;
	$S_2 \leftarrow \textsf{initialise}(w-h) + \overrightarrow{1} + S[h]+h-1$\;
	$S_3 \leftarrow S^r + w- 1$\;
	$\outputleft \leftarrow \textsf{join}(\textsf{join}(S_1, S_2), S_3)$\;
	
	 \caption{Compute the shortest path passing through the left border}\label{alg:left}
	\end{algorithm}\label{alg:outleft}

Now, equations~\eqref{eqn:left} and~\eqref{eqn:top} can be rephrased as \cref{alg:left,alg:top}, respectively.
The final step of the algorithm is to compute the \outborder as the minimum of $\outputtop[i]$ and $\outputleft[i]$ for each index $i$.
This is performed in linear time per block by Chen and Chao~\cite{CC-RLEedit:2013}. In \cref{sec:newalgo} we 
will design a new implementation of both algorithms and a subtle amortised argument for this final step.
The latter is based on the fundamental property of the values in an \outborder summarised by \cref{lem:cross}.

\begin{lemma}[{\cite[Lemma 7]{CC-RLEedit:2013}}]\label{lem:cross}
	If there exists an $i$ such that $\outputtop[i]\leq\outputleft[i]$, then $\outputtop[j]\leq\outputleft[j]$ for all $j\geq i$.
\end{lemma}

%\begin{algorithm}[t]
%
%	$S\; \leftarrow \textsf{SWM}_h(\inputleft[1,h])$\;
%	$S_1 \leftarrow \textsf{split}(S,1,h) + \overrightarrow{1} -1$\;
%	$S_2 \leftarrow \textsf{initialise}(w-h) + \overrightarrow{1} + S[h]+h-1$\;
%	$S_3 \leftarrow  \textsf{split}(S,h+1,2h-1) + w- 1$\;
%	$\outputleft \leftarrow \textsf{Concat}(S_1, S_2, S_3)$\;
%	
%	 \caption{Compute the shortest path passing through the left border}\label{alg:left}
%	\end{algorithm}\label{alg:outleft}

%\begin{algorithm}[t]
%
%	$S\; \leftarrow \textsf{SWM}_h(\inputtop[1,w])$\;
%	$S_1 \leftarrow  \textsf{split}(S, 1, w) + h - 1$\;
%	$S_2 \leftarrow \textsf{split}(S,w+1,w+h-1) - \overrightarrow{1} + w+ h - 1$\;
%	$\outputtop \leftarrow \textsf{Concat}(S_1, S_2)$\;
%	
%	\caption{Compute the shortest path passing through the top border}\label{alg:top}
%\end{algorithm}

\begin{algorithm}[t]

  $S\; \leftarrow \textsf{SWM}_h(\inputtop[1,w])$\;
  $S^\ell, S^r \leftarrow  \textsf{split}(S, w)$\;
  $S_1 \leftarrow  S^\ell + h - 1$\;
  $S_2 \leftarrow S^r - \overrightarrow{1} + w+ h - 1$\;
  $\outputtop \leftarrow \textsf{join}(S_1, S_2)$\;
  
  \caption{Compute the shortest path passing through the top border}\label{alg:top}
\end{algorithm}

Before we go on to explain how we speed up the task of deriving the borders of blocks, it is worth exploring for a moment why we cannot simply apply, perhaps with some small modifications, the known $O(mn\log(mn))$ time solution for LCS on RLE strings~\cite{ALS:1999}.  The key obstacle comes from the different nature of optimal paths in the dynamic programming table of the LCS and edit distance problems.

\begin{figure}[b!]
	 \begin{minipage}{0.46\textwidth}
		\centering
	\begin{tikzcd}[column sep=4pt, row sep=6pt,nodes={inner sep=0pt,align=center,
	text width={width("20")}},nodes in empty cells,
execute at end picture={
	\foreach \x in {2,5,11}
	{
		\draw 
		([yshift=-3,xshift=-.125\pgflinewidth]\tikzcdmatrixname-\x-1.south west) --   
		([yshift=-3,xshift=-.125\pgflinewidth]\tikzcdmatrixname-\x-11.south east);
	}
	\foreach \y in {3}
	{
		\draw 
		([yshift=.5\pgflinewidth]\tikzcdmatrixname-1-\y.north west) -- 
		([yshift=.5\pgflinewidth]\tikzcdmatrixname-14-\y.south west);
}}]
&&\texttt{a}&\texttt{a}&\texttt{a}&\texttt{a}&\texttt{a}&\texttt{a}&\texttt{a}&\texttt{a}&\texttt{a}\\
&\ar[dr]0&1&2&3&4&5&6&7&8&9\\
\texttt{a}&1&\ar[dr]0&1&2&3&4&5&6&7&8\\
\texttt{a}&2&1&\ar[dr]0&1&2&3&4&5&6&7\\
\texttt{a}&3&2&1&\ar[d]\ar[dr]0&1&2&3&4&5&6\\
\texttt{b}&4&3&2&\ar[d]1&\ar[dr]1&2&3&4&5&6\\
\texttt{b}&5&4&3&\ar[d]2&2&\ar[dr]2&3&4&5&6\\
\texttt{b}&6&5&4&\ar[dr]3&3&3&\ar[d]3&4&5&6\\
\texttt{b}&7&6&5&4&\ar[dr]4&4&\ar[d]4&4&5&6\\
\texttt{b}&8&7&6&5&5&\ar[dr]5&\ar[d]5&5&5&6\\
\texttt{b}&9&8&7&6&6&6&\ar[dr]6&6&6&6\\
\texttt{a}&10&9&8&7&6&6&6&\ar[dr]6&6&6\\
\texttt{a}&11&10&9&8&7&6&6&6&\ar[dr]6&6\\
\texttt{a}&12&11&10&9&8&7&6&6&6&6\\
\end{tikzcd}
	\caption{Edit distance with forced turn in mismatch block}\label{fig:ed-table}
	  \end{minipage}\hfill
   \begin{minipage}{0.46\textwidth}
  	\centering
	\begin{tikzcd}[column sep=4pt, row sep=6pt,nodes={inner sep=0pt,align=center,
	text width={width("20")}},nodes in empty cells,
execute at end picture={
	\foreach \x in {2,5,11}
	{
		\draw 
		([yshift=-3,xshift=-.125\pgflinewidth]\tikzcdmatrixname-\x-1.south west) --   
		([yshift=-3,xshift=-.125\pgflinewidth]\tikzcdmatrixname-\x-11.south east);
	}
	\foreach \y in {3}
	{
		\draw 
		([yshift=.5\pgflinewidth]\tikzcdmatrixname-1-\y.north west) -- 
		([yshift=.5\pgflinewidth]\tikzcdmatrixname-14-\y.south west);
}}]
&&\texttt{a}&\texttt{a}&\texttt{a}&\texttt{a}&\texttt{a}&\texttt{a}&\texttt{a}&\texttt{a}&\texttt{a}\\
&0\ar[r]&0\ar[r]&0\ar[r]&0\ar[dr]&0&0&0&0&0&0\\
\texttt{a}&0&1&1&1&1\ar[dr]&1&1&1&1&1\\
\texttt{a}&0&1&2&2&2&2\ar[dr]&2&2&2&2\\
\texttt{a}&0&1&2&3&3&3&3\ar[d]&3&3&3\\
\texttt{b}&0&1&2&3&3&3&3\ar[d]&3&3&3\\
\texttt{b}&0&1&2&3&3&3&3\ar[d]&3&3&3\\
\texttt{b}&0&1&2&3&3&3&3\ar[d]&3&3&3\\
\texttt{b}&0&1&2&3&3&3&3\ar[d]&3&3&3\\
\texttt{b}&0&1&2&3&3&3&3\ar[d]&3&3&3\\
\texttt{b}&0&1&2&3&3&3&3\ar[dr]&3&3&3\\
\texttt{a}&0&1&2&3&4&4&4&4\ar[dr]&4&4\\
\texttt{a}&0&1&2&3&4&5&5&5&5\ar[dr]&5\\
\texttt{a}&0&1&2&3&4&5&6&6&6&6\\
\end{tikzcd}
  		\caption{LCS with optimal vertical path through mismatch block}\label{fig:lcs-table}
  \end{minipage}
	
\end{figure}

%\begin{figure}[t]
%		\centering
%	\begin{tikzcd}[column sep=3pt, row sep=5pt,nodes={inner sep=0pt,align=center,
%		text width={width("20")}},nodes in empty cells,
%	execute at end picture={
%		\foreach \x in {2,5,11}
%		{
%			\draw 
%			([yshift=-3,xshift=-.125\pgflinewidth]\tikzcdmatrixname-\x-1.south west) --   
%			([yshift=-3,xshift=-.125\pgflinewidth]\tikzcdmatrixname-\x-11.south east);
%		}
%		\foreach \y in {3}
%		{
%			\draw 
%			([yshift=.5\pgflinewidth]\tikzcdmatrixname-1-\y.north west) -- 
%			([yshift=.5\pgflinewidth]\tikzcdmatrixname-14-\y.south west);
%	}}]
%    &&\texttt{a}&\texttt{a}&\texttt{a}&\texttt{a}&\texttt{a}&\texttt{a}&\texttt{a}&\texttt{a}&\texttt{a}\\
%    &\ar[dr]0&1&2&3&4&5&6&7&8&9\\
%\texttt{a}&1&\ar[dr]0&1&2&3&4&5&6&7&8\\
%\texttt{a}&2&1&\ar[dr]0&1&2&3&4&5&6&7\\
%\texttt{a}&3&2&1&\ar[d]\ar[dr]0&1&2&3&4&5&6\\
%\texttt{b}&4&3&2&\ar[d]1&\ar[dr]1&2&3&4&5&6\\
%\texttt{b}&5&4&3&\ar[d]2&2&\ar[dr]2&3&4&5&6\\
%\texttt{b}&6&5&4&\ar[dr]3&3&3&\ar[d]3&4&5&6\\
%\texttt{b}&7&6&5&4&\ar[dr]4&4&\ar[d]4&4&5&6\\
%\texttt{b}&8&7&6&5&5&\ar[dr]5&\ar[d]5&5&5&6\\
%\texttt{b}&9&8&7&6&6&6&\ar[dr]6&6&6&6\\
%\texttt{a}&10&9&8&7&6&6&6&\ar[dr]6&6&6\\
%\texttt{a}&11&10&9&8&7&6&6&6&\ar[dr]6&6\\
%\texttt{a}&12&11&10&9&8&7&6&6&6&6\\
%  \end{tikzcd}
%\end{figure}

For the LCS problem on RLE strings Apostolico et al.~\cite{ALS:1999} introduced two important concepts. The first is \emph{forced paths} and the second \emph{corner paths}.  They say that a path beginning at the upper-left corner of a match block is \emph{forced} if it traverses the block by strictly diagonal moves and, whenever the right (respectively,
lower) side of an intermediate match block is reached, proceeds to the next
match block by a straight horizontal (respectively, vertical) line through
the mismatch blocks in between. A \emph{corner} path is one that enters match blocks in the top left corner and exits only through the bottom right corner. They show that there is always a longest common subsequence between two strings corresponding to the concatenation of subpaths of corner and forced paths. This fact greatly reduces the number of different paths that have to be considered and hence the complexity of solving the overall LCS problem. However for the edit distance problem this property of forced paths no longer holds. Figures~\ref{fig:ed-table} and \ref{fig:lcs-table} show an example of this key difference between optimal paths under edit distance and LCS. In Figure~\ref{fig:ed-table} we can see that there is no optimal vertical (or horizontal) path through the mismatch block. By contrast, there is indeed an optimal vertical path for the LCS problem as illustrated by Figure~\ref{fig:lcs-table}.

In order to speed up edit distance computation on RLE strings we  introduce a new data structure for \inborders and \outborders. This will allow us to derive the values of \outborders from their respective \inborders in amortised logarithmic time per border, rather than the previous linear time.  The rest of the paper is devoted to this task.

% !TEX root = main.tex
\section{Efficient Manipulation of Piecewise-Linear Functions}\label{sec:newalgo}

In this section, we describe the data structure we will use to represent \inborders and \outborders in the dynamic programming table.  We will then show how the operations from \cref{definitions} can be implemented efficiently using this data structure.

Recall that \emph{a piecewise linear function} is a real-valued function $F$ 
whose domain $\dom(F)$ is a closed interval that can be represented as a union of closed intervals $\dom(F)=\bigcup_{j=1}^k I_j$
such that $F$ restricted to $I_j$ is an affine function ($g_jx+h_j$ for some coefficients $g_j$ and $h_j$).  The input and \outborders as defined in \cref{sec:previous} are by definition piecewise linear.

In this section, we impose a few further restrictions:
\begin{itemize}[itemsep=0ex, parsep=1pt, topsep=1pt]
	\item For each integer $x\in \dom(F)$, the value $F(x)$ is also an integer.
	\item The gradient $g_j$ of each $F|_{I_j}$ is $-1$, $0$, or $1$.
	\item The endpoints of $\dom(F)$ are integers or half-integers.
\end{itemize}
The graph of a piecewise linear function $F$ is a simple polygonal curve, and thus it can be interpreted as a sequence of \emph{turning points}
connected by straight-line \emph{segments}. 
Due to the restrictions imposed on $F$, each turning point has integer or half-integer coordinates.
We represent such a function $F$ as a sequence of segments stored in an annotated balanced binary search tree,
where each segment explicitly keeps the coordinates of its endpoints.\footnote{Note that the coordinates of each internal turning point are stored with both incident segments.}

We first provide a simple implementation of curves supporting a few basic operations,
and then we gradually augment it to handle more complicated operations.
We conclude with an amortised running time analysis.

\subsection{Basic Operations}

Our first implementation just stores the corresponding segment for each node $v$:

\begin{description}[itemsep=0ex, parsep=1pt, topsep=1pt]
	\item[{\bf $(x_\ell, y_\ell)$:}] The coordinates of the left endpoint of the segment corresponding to $v$.
	\item[{\bf $(x_r, y_r)$:}] The coordinates of the right endpoint of the segment corresponding to $v$.
\end{description}

\noindent
Nevertheless, we are already able to implement some operations useful in \cref{alg:left,alg:top}.

\subparagraph*{Create}   The \textsf{create} operation produces a function $F$ whose graph consists of just one segment $S$
with given endpoints $(x_\ell,y_\ell)$ and $(x_r,y_r)$. This enables us to implement the whole of line 4 of \cref{alg:left} in worst-case constant time.

\subparagraph*{Join} The \textsf{join} operation takes two functions, $F_L$ and $F_R$
with domains $\dom(F_L)=[x_L, x_M]$ and $\dom(F_R)=[x_M,x_R]$, respectively, and with a common endpoint $F_L(x_M) = F_R(x_M)$.
It returns a function $F$ with $\dom(F)=[x_L,x_R]$ such that $F_L= F|_{[x_L,x_M]}$ and $F_R = F|_{[x_M,x_R]}$.
To implement this operation, we first join the two balanced binary search trees.
If the rightmost segment of $F_L$ has the same gradient as the leftmost segment of $F_R$, we also join these segments.
The resulting tree represents $F$. The worst-case running time is logarithmic.

\subparagraph*{Split} The \textsf{split} operation takes a function $F$ with $\dom(F)=[x_L,x_R]$ and a value $x_M\in \dom(F)$.
It returns two functions $F_L = F|_{[x_L,x_M]}$ and $F_R = F|_{[x_M,x_R]}$.
To implement it, we first descend the binary search tree to find a segment $S$ with $x_M\in \dom(S)$.
If $x_M$ lies in the interior of $\dom(S)$, we split this segment into two.
Next, we split the binary search tree to separate the segments to the left of $x_M$ from the segments to the right of $x_M$.
The resulting two trees represent $F_L$ and $F_R$, respectively.
The worst-case running time is logarithmic.

\subparagraph*{Combine} 
The \textsf{combine} operation takes two functions $F_1$ and $F_2$ over the same domain $\dom(F_1) = \dom(F_2)=[x_L,x_R]$ and returns their pointwise minimum:
a function $F$ with $\dom(F)=[x_L,x_R]$ such that $F(x) = \min(F_1(x), F_2(x))$ for $x\in\dom(F)$.
We assume that there exists $x_M\in [x_L,x_R]$ such that $F_1(x) > F_2(x)$ if $x < x_M$ and $F_1(x)\le F_2(x)$ if $x \geq x_M$.

If $F_1(x_L) \le F_2(x_L)$, then $x_M = x_L$. Hence, we return $F=F_1$ and discard $F_2$.
Similarly, if $F_1(x_R) > F_2(x_R)$, then $x_M = x_R$. Hence, we return $F = F_2$ and discard $F_1$.

Otherwise, we are guaranteed that $F_1(x_M)=F_2(x_M)$. Our first task is to find $x_M$.
For this, we locate segments $S_1$ of $F_1$ and $S_2$ of $F_2$ such that $x_M \in \dom(S_1)\cap \dom(S_2)$.

We observe that $S_1$ corresponds to the leftmost node $v$ in the BST of $F_1$ such that
 $v.y_r = F_1(v.x_r) \le F_2(v.x_r)$.
Hence, we perform a left-to-right in-order traversal of the BST to find $S_1$.
For each visited node $v$, we evaluate $F_2(v.x_r)$ by descending the BST of $F_2$ to find a segment whose domain contains $v.x_r$.
Symmetrically, $S_2$ corresponds to the rightmost node $v$ in the BST of $F_2$ such that
$F_1(v.x_\ell) > F_2(v.x_\ell)=v.y_\ell$, so we find $S_2$ by performing a right-to-left in-order traversal of the BST.

Next, we note that  $(x_M, F_1(x_M))=(x_M, F_2(x_M))$ is the leftmost common point of $S_1$ and $S_2$.
Hence, we can now compute $x_M$ easily (the restrictions on $F_1$ and $F_2$ guarantee that it is an integer or a half-integer).
Finally, we \textsf{split} both $F_1$ and $F_2$ at $x_M$, discard $F_1|_{[x_L,x_M]}$ and $F_2|_{[x_M,x_R]}$,
\textsf{join} $F_2|_{[x_L,x_M]}$ with $F_1|_{[x_M,x_R]}$, and return the resulting function as $F$.

As far as the running time is concerned, the cost is logarithmic for each discarded segment.
We can now also implement the final \textsf{combine} step that produces our representation of the \outborder from the outputs of \cref{alg:left,alg:top} by finding the minimum at each index.

\subsection{Shifts}\label{sec:shift}
Next, we extend our data structure to implement the \textsf{shift} operation
which moves the whole function by a given vector. It is useful in \cref{alg:left,alg:top} for altering $S^\ell$ and $S^r$.

Formally, given a function $F$ with $\dom(F)=[x_L,x_R]$ and a vector $(\Delta_x,\Delta_y)$,
we transform $F$ into $F'$ such that $F'(x) = F(x-\Delta_x)+\Delta_y$ for each $x\in \dom(F')=[x_L+\Delta_x, x_R+\Delta_x]$.

This update is performed using a technique known as \emph{lazy propagation}.
We augment each node $v$ with the following extra field:

\begin{description}[itemsep=0ex, parsep=1pt, topsep=1pt]
	\item[{\bf $(\delta_x, \delta_y)$:}] A deferred shift to be propagated within the subtree of $v$.
\end{description}

\noindent
This change is then lazily propagated as further operations are executed. Here, we rely on a key structural property of BST operations:
\begin{observation}\label{obs:bst}
	The execution of every BST operation can be extended (at the cost of an extra multiplicative constant in the running time) with a sequence of node \emph{activations} and \emph{deactivations} such that:
	\begin{itemize}[itemsep=0ex, parsep=1pt, topsep=1pt]
		\item a node $v$ is accessed only when it is active and has no active descendant,
		\item when $v$ is active, then all its ancestors are active,
		\item no node is active at the beginning and the end of the execution.
	\end{itemize}
\end{observation}

The idea behind lazy propagation is that the deferred updates stored at a node $v$ are propagated when $v$ is activated.
This way, every active node has no delayed updates pending. Hence, from the perspective of any other operation, the effect is the same as if we have meticulously modified every node for each update.

The shift propagation is very simple: when a node $v$ receives a request for a shift by $(\Delta_x,\Delta_y)$,
then we just add $(\Delta_x,\Delta_y)$ to the delayed shift $(v.\delta_x, v.\delta_y)$ stored at $v$.
Upon activation of $v$, we propagate $(v.\delta_x, v.\delta_y)$ to the children of $v$,
add $(v.\delta_x, v.\delta_y)$ to both $(x_\ell, y_\ell)$ and $(x_r,y_r)$, and reset $(v.\delta_x, v.\delta_y):=(0,0)$.
To implement the \emph{shift} operation, we just send a request for a shift by $(\Delta_x,\Delta_y)$ to the root node $r$.

The worst-case running time of a \textsf{shift} is constant, and the extra cost of propagation does not increase the asymptotic running time
of the remaining operations.

\subsection{Gradient Changes}\label{sec:grad}

The \textsf{gradient change} operation takes a function $F$ and a coefficient $\Delta_g$, and it transforms $F$ into $F'$ 
such that $F'(x) = F(x)+\Delta_g\cdot x$ for each $x\in \dom(F')=\dom(F)$. This operation is needed in both \cref{alg:left,alg:top}
to transform $S^r$ and $S^\ell$, respectively.

We first note that the constraints imposed on the gradients of functions $F$ and $F'$ yield that $\Delta_g = -1$,
$F$ is non-decreasing, and $F'$ is non-increasing, or $\Delta_g = 1$, $F$ is non-increasing, and $F'$ is non-decreasing.
However, these limitations only become relevant in \cref{sec:swm}.

To implement \textsf{gradient change}, we just add another field to each node $v$:

\begin{description}[itemsep=0ex, parsep=1pt, topsep=1pt]
	\item[{\bf $\delta_g$:}] A deferred gradient change to be propagated within the subtree of $v$.
\end{description}

\noindent
We now have two types of lazily propagated updates: \textsf{shift} and \textsf{gradient change}.
These two operations do \emph{not} commute, so we need decide how to interpret the two kinds of deferred updates stored at a node $v$.
We shall assume that the gradient change by $\delta_g$ is to be performed \emph{before} the shift by $(\delta_x,\delta_y)$. 

Thus, while shift propagation is implemented as in \cref{sec:shift},
adding $\Delta_g$ to $v.\delta_g$ is insufficient when a node $v$ receives a request to change gradient by $\Delta_g$:
we also need to add $\Delta_g \cdot \delta_x$ to $\delta_y$. This approach is correct since a shift by $(\Delta_x, \Delta_y)$ followed by a gradient change by $\Delta_g$ is equivalent to a gradient change by $\Delta_g$ followed by a shift by $(\Delta_x, \Delta_y+\Delta_g \cdot \Delta_x)$. 

Upon activation of $v$, we first apply the deferred gradient change: we propagate it to the children of $v$,
increase $v.y_\ell$ by $ v.\delta_g\cdot v.x_\ell$ and  $v.y_r$ by $ v.\delta_g\cdot v.x_r$, and reset $v.\delta_g = 0$.
Then, we handle the deferred shift as in \cref{sec:shift}.

Finally, we note that to implement the \emph{gradient change} operation, we just send a request for a gradient change by $\Delta_g$ to the root node $r$.
The worst-case running time is constant.

\subsection{Sliding Window Minima}\label{sec:swm}

We can finally show how to implement the \SWM operation efficiently on our data structure. This is the most involved of the operations we will need.
The \SWM operation given a function $F$ with $\dom(F)=[x_L,x_R]$ and a \emph{window width} $t$,
returns a function $F'$ with $\dom(F')=[x_L, x_R+t]$ such that $F'(x)=\min\{F(x') : x' \in [x,x-t]\cap \dom(F)\}$.

\paragraph*{Combinatorial Properties}
We begin by observing that the \SWM operation is composable.

\begin{observation}\label{obs:compos}
	Every function $F$ and positive window widths $t,t'$ satisfy
	$\SWM(\SWM(F,t),t')=\SWM(F,t+t')$.
\end{observation}
% \begin{proof}
% 	\begin{eqnarray*}
% 	(S^{(h)})^{(t)}[i] &=&\min\{S^{(h)}[j]|i-t\leq j\leq i, 0\leq j\leq n+h\}\\
% 	&=&\min\{\min\{S[k]|j-h\leq k\leq j, 0\leq k\leq n\}|i-t\leq j\leq i, 0\leq j\leq n+h\}\\
% 	&=&\min\{S[k]|i-(t+h)\leq k\leq i, 0\leq k\leq n\}\\
% 	&=& S^{(t+h)}[i]
% \end{eqnarray*}
% \end{proof}
Hence, instead of applying $\SWM(\cdot, t)$ for an integer width $t$,
we may equivalently apply the $\SWM(\cdot, 1)$ operation $t$ times.
The key property of width $1$ is that the changes to the transformed function are very local.
The structure of these modifications can be described in terms of \emph{types} of turning points.
We classify internal turning points by the gradients (\textbf{I}ncreasing, \textbf{F}lat, or \textbf{D}ecreasing) of the 
incident segments; see \cref{table:one}, where we also analyse how a function changes in the vicinity of each turning point
subject to $\SWM(\cdot, 1)$.
\begin{table}[thb]
	\centering
	\begin{tabular}{c|c|c|c|c|c}
		Type DI & Type ID & Type IF & Type FD & Type FI & Type DF\\
		\hline
		\tikz{
			\draw[fill] (0,0.5) -- (0.5, 0) circle[radius=1pt] -- (1, 0.5);
			\path(0,0.75);
			\begin{scope}[yshift=-.8cm]
			\draw[dotted, fill=black!50] (0,0.5) -- (0.5, 0) circle[radius=1pt] -- (1, 0.5);
			\draw[fill] (0,0.5) -- (0.5, 0) circle[radius=1pt] -- (1, 0) circle[radius=1pt] -- (1.5, 0.5);
			\end{scope}
			}
		&
		\tikz{
			\draw[fill] (0,0) -- (0.5, 0.5) circle[radius=1pt] -- (1, 0);
			\path(0,0.75);
			\begin{scope}[yshift=-.8cm]
			\draw[dotted, fill=black!50] (0,0) -- (0.5, 0.5) circle[radius=1pt] -- (1, 0);
			\draw[fill] (0.5,0) -- (0.75, 0.25) circle[radius=1pt] -- (1, 0);
			\end{scope}			
			}
		&    
		\tikz{
			\draw[fill] (0,0) -- (0.5, 0.5) circle[radius=1pt] -- (1.5, 0.5);
			\path(0,0.75);
			\begin{scope}[yshift=-.8cm]
			\draw[dotted, fill=black!50] (0,0) -- (0.5, 0.5) circle[radius=1pt] -- (1.5, 0.5);
			\draw[fill] (0.5,0) -- (1, 0.5) circle[radius=1pt] -- (1.5, .5);
			\end{scope}			
		}
		&   
		\tikz{
			\draw[fill] (0,0.5) -- (0.5, 0.5) circle[radius=1pt] -- (1, 0);
			\path(0,0.75);
			\begin{scope}[yshift=-.8cm]
			\draw[dotted, fill=black!50] (0,0.5) -- (0.5, 0.5) circle[radius=1pt] -- (1, 0);
			\draw[fill] (0,0.5) -- (0.5, 0.5) circle[radius=1pt] -- (1, 0);
			\end{scope}	
			} 
		&
		\tikz{
			\draw[fill] (0,0) -- (0.5, 0) circle[radius=1pt] -- (1, 0.5);
			\path(0,0.75);
			\begin{scope}[yshift=-.8cm]
			\draw[dotted, fill=black!50] (0,0) -- (0.5, 0) circle[radius=1pt] -- (1, 0.5);
			\draw[fill] (0,0) -- (1, 0) circle[radius=1pt] -- (1.5, .5);
			\end{scope}	
			} 
		&
		\tikz{
			\draw[fill] (0,0.5) -- (0.5, 0) circle[radius=1pt] -- (1, 0);
			\path(0,0.75);
			\begin{scope}[yshift=-.8cm]
			\draw[dotted, fill=black!50] (0,0.5) -- (0.5, 0) circle[radius=1pt] -- (1, 0);
			\draw[fill] (0,0.5) -- (0.5, 0) circle[radius=1pt] -- (1, 0);
			\end{scope}	
			}
	\end{tabular}
	\caption{Types of internal turning points and their behaviour subject to the \SWM operation.}
	\label{table:one}
\end{table}

\begin{itemize}[itemsep=0ex, parsep=1pt, topsep=1pt]
		\item A point of type FD or DF remains intact.
		\item A point of type FI or IF is shifted by $(1,0)$.
		\item A point of type ID is shifted by $(0.5, -0.5)$.
		\item A point of type DI transformed into a point of type DF and a point of type FI,
		and the latter is shifted by $(1,0)$.
	\end{itemize}

Note that the behaviour of DI points is unlike that of other types. However, this discrepancy disappears if we replace
every DI point with two coinciding points of types DF and FI, respectively, with an artificial length-0 segment in between.
Hence, whenever a new internal turning point is created (which happens only within the \textsf{join} operation),
if the turning point would be of type DI, we pre-emptively replace it by two coinciding points of type DF and FI, respectively.
Note that the resulting length-0 segment never changes its gradient since \textsf{gradient change} is allowed only on a monotone function.
However, when an incident segment is modified, we may need to remove the length-0 segment. This process cannot cascade, though, causing another
length zero segment to be removed.

\begin{table}[b!]
	\centering
	\begin{tabular}{c|c|c|c|c|c}
		Type -I & Type -F & Type -D & Type I- & Type F- & Type D-\\
		\hline
		\tikz{
			\draw[fill] (0.5, 0) circle[radius=1pt] -- (1, 0.5);
			\path(0.5,0.75);
			\begin{scope}[yshift=-.8cm]
			\draw[dotted, fill=black!50] (0.5, 0) circle[radius=1pt] -- (1, 0.5);
			\draw[fill] (0.5, 0) circle[radius=1pt] -- (1, 0) circle[radius=1pt] -- (1.5, 0.5);
			\end{scope}
			}
		&
		\tikz{
			\draw[fill] (0.5, 0.25) circle[radius=1pt] -- (1, 0.25);
			\path(0.5,0.75);
			\begin{scope}[yshift=-.8cm]
			\draw[dotted, fill=black!50] (0.5, 0.25) circle[radius=1pt] -- (1, 0.25);
			\draw[fill] (0.5, 0.25) circle[radius=1pt] -- (1, 0.25);
			\end{scope}			
			}
		&    
		\tikz{
			\draw[fill] (0.5, 0.5) circle[radius=1pt] -- (1, 0);
			\path(0.5,0.75);
			\begin{scope}[yshift=-.8cm]
			\draw[dotted, fill=black!50] (0.5, 0.5) circle[radius=1pt] -- (1, 0);
			\draw[fill] (0.5, 0.5) circle[radius=1pt] -- (1, 0);
			\end{scope}			
		}
		&   
		\tikz{
			\draw[fill] (0, 0) -- (0.5, 0.5) circle[radius=1pt];
			\path(0.5,0.75);
			\begin{scope}[yshift=-.8cm]
			\draw[dotted, fill=black!50] (0, 0) -- (0.5, 0.5) circle[radius=1pt];
			\draw[fill] (0.5, 0) -- (1, 0.5) circle[radius=1pt];
			\end{scope}	
			} 
		&
		\tikz{
			\draw[fill] (0, 0.25) -- (0.5, 0.25) circle[radius=1pt];
			\path(0,0.75);
			\begin{scope}[yshift=-.8cm]
			\draw[dotted, fill=black!50] (0, 0.25) -- (0.5, 0.25) circle[radius=1pt];
			\draw[fill] (0, 0.25) -- (1, 0.25) circle[radius=1pt];
			\end{scope}	
			} 
		&
		\tikz{
			\draw[fill] (0, 0.5) -- (0.5, 0) circle[radius=1pt];
			\path(0,0.75);
			\begin{scope}[yshift=-.8cm]
			\draw[dotted, fill=black!50] (0, 0.5) -- (0.5, 0) circle[radius=1pt];
			\draw[fill] (0, 0.5) -- (0.5, 0) circle[radius=1pt] -- (1, 0)  circle[radius=1pt];
			\end{scope}	
			}
	\end{tabular}
	\caption{Types of endpoints and their behaviour subject to the \SWM operation.}
	\label{table:two}
\end{table}

Next, we analyse in \cref{table:two} how the $\SWM(F,1)$ operation affects the endpoints of the graph of $F$.
In most cases, the left endpoint stays intact and the right endpoint is shifted by $(0,1)$.
The only exceptions are endpoints of type -I and D-, which exhibit similar behaviour to the internal turning points of type DI.
Moreover, this discrepancy also disappears if we introduce artificial flat segments of length 0. Hence,
we replace a point of type -I with two points of type -F and FI, respectively,
and a point of type D- with two points of type DF and F-, respectively.
However, this time the replacement is \emph{not} pre-emptive: we perform it as the first step in the implementation of the \SWM operation.
This is possible because there are just two endpoints, while the number of internal turning points of type DI could be large.
Our gain, on the other hand, is that we avoid length-0 segments changing their gradients. 

With the artificial length-0 segments in place, it is now true that the effect of the \SWM operation on each turning point
can be described as a shift depending only on the type of the point.
As a result of these shifts, some segments may disappear as their length reaches 0; in this case, we say that a segment \emph{collapses}.
Only segments of three kinds may collapse:
\begin{itemize}[itemsep=0ex, parsep=1pt, topsep=1pt]
	\item a segment between a point of type ID and point of type DF;
	\item a segment between a point of type IF and point of type FD;
	\item a segment between a point of type FI and point of type ID;
\end{itemize}
When a segment collapses, it is removed and the two incident turning points are merged.\footnote{Two adjacent segments may collapse simultaneously. In that special case, three subsequent points, of type FI, ID, and DF, respectively, need to be deleted.} 
Each segment of the three affected kinds has the \emph{collapse time},
defined as the smallest $t$ for which $\SWM(\cdot, t)$ makes it collapse (assuming no interaction from incident segments)
equal the the Manhattan distance between its endpoints.
Note that due to the restrictions on the piecewise linear functions considered in this section, the collapse time is always an integer.

\paragraph*{Implementation}
To implement the \SWM operation, we augment each node $v$ with the following fields:

\begin{description}[itemsep=0ex, parsep=1pt, topsep=1pt]
	\item[{\bf $\textrm{type}_\ell, \textrm{type}_r$:}] The types of the turning points joined by the segment corresponding to $v$.
	\item[{\bf $\delta_t$:}] The amount of a deferred \SWM to be propagated within the subtree of $v$.
	\item[{\bf $t_{\min}$:}] The minimum collapse time among the segments in the subtree of $v$.
\end{description}

Note that the type of each internal turning point is stored twice. Hence, whenever a node type changes,
this fact needs to be reflected at both incident segments (and we need to reach the corresponding nodes by descending the BST;
shortcuts would violate \cref{obs:bst}).

The field  $v.t_{\min}$ is of a kind we have not encountered yet:
 its value depends on the corresponding values for the children of $v$ and on other fields at $v$.
It is brought up to date whenever $v$ is deactivated (so that it can be accessed only when $v$ is inactive).
We shall assume that its value already reflects the deferred updates stored at $v$.
The procedure of recomputing  $v.t_{\min}$ is simple: we determine the collapse time of the segment represented by $v$
(which is infinite or equal to $|v.x_r-v.x_\ell|+|v.y_r-v.y_\ell|$ depending on the types of the incident turning points),
and take the minimum of this value and $u.t_{\min}$ for every child $u$ of $v$.
Since $v$ has no deferred changes when it is deactivated, the resulting minimum is $v.t_{\min}$.

\subparagraph*{Propagation}
The main structural modification to the lazy propagation procedures is that we maintain an additional invariant that no deferred changes
are stored on the leftmost and on the rightmost path of the BSTs representing every function $F$.
To maintain this invariant, immediately after lazily updating of the whole $F$ (sending a request to the root node $r$), we descend to the leftmost and to the rightmost segment $F$; this increases the cost of \textsf{shift} and \textsf{gradient change} to logarithmic.
Note that the \textsf{split} operation must anyway visit the nodes representing the new boundary segments (to update the types of new endpoints).
Moreover, if a path from the root to a given node $v$ contains no deferred updates, then this is still true after any rebalancing of the BST
(as only active nodes get rotated).

Concerning the lazy \SWM propagation, we explicitly forbid requesting for \SWM with window width exceeding $r.t_{\min}$,
because collapsed segments need to be removed before we proceed further. Also, the window widths (and hence the values $\delta_t$) are always non-negative.

We have three kinds of deferred updates now:
\SWM, \textsf{gradient change}, and \textsf{shift}. We fix the semantics of the fields $\delta_t$, $\delta_g$, and $(\delta_x,\delta_y)$ so that 
an \SWM of width $\delta_t$ is performed first, a \textsf{gradient change} by $\delta_g$ second, and  a shift by $(\delta_x,\delta_y)$ last.
The requests for a shift and for a gradient change are still implemented as in \cref{sec:grad};
note that these updates do not affect the collapse times (the three segment kinds with finite collapse times cannot appear in monotone functions).
On the other hand, the request for an \SWM with a window width $\Delta_t$ requires more care.
We clearly need to increase $\delta_t$ by $\Delta_t$ and decrease $t_{\min}$ by the same amount.
The aforementioned steps suffice if $\delta_g = 0$.
Otherwise, we note that the turning points in the subtree of $v$ are all of types DF and FD or all of types IF and FI.
(Observe that there are no deferred changes in the proper ancestors of $v$ and that $v$ is not on the leftmost or rightmost path in this case.)
We can easily distinguish the two cases by analysing the endpoints
of the segment corresponding to $v$. Moreover, the \SWM operation is void in the first of these cases, and in the second one it reduces to a shift by $(\Delta_t, 0)$. Hence, we shall implement it this way rather than by modifying $v.\delta_t$.

The propagation itself is relatively easy: upon activation of a node $v$,
we first propagate the \SWM operation to the children of $v$, update the endpoints of the segment corresponding to $v$ (according to \cref{table:one,table:two}, with the shift multiplied by $\delta_t$), and finally reset $\delta_t = 0$.
Then, we propagate the gradient change and the shift. This is implemented as in \cref{sec:grad} except that the gradient change
now affects not only the coordinates but also the types of the segment's endpoints.

\subparagraph*{\SWM Procedure}
To implement the \SWM procedure itself, we first check the endpoint types and perform appropriate substitutions for endpoints of type -I and D-. Next, we would like to lazily apply the \SWM operation with window width $t$ to the root $r$.
However, this could result in negative collapse time $r.t_{\min}$,
so instead we perform \SWM gradually based on \cref{obs:compos}. 
If $r.t_{\min} < t$, we make a request for \SWM with window width just $r.t_{\min}$, leaving the remaining
quantity $t - r.t_{\min}$ for later on. This already results in $r.t_{\min}=0$, which indicates that there is a collapsed segment.
We descend the tree to find such a collapsed segment (activating nodes on the way there and deactivating them on the way back),
and take care of this segment appropriately (this may affect neighbouring segments as well).
We repeat the process as long as $r.t_{\min}=0$. Once this value is positive again, we are ready to proceed with further application of \SWM.

As far as the the running time is concerned, the cost of \SWM consists of a logarithmic term for visiting the endpoints and further logarithmic terms for each collapsed segment. 

\subsection{Complexity Analysis}

We complete this section by showing that the aforementioned operations run in \emph{amortised} logarithmic time.

\begin{lemma}
\label{lem:overall}
	A sequence of $k$ operations on piecewise linear functions takes $\Oh(k \log k)$ time.
\end{lemma}
\begin{proof}
	Our potential is $\log k$ multiplied by the total number of turning points in all the stored functions.
	First, we observe that this potential grows by $\Oh(\log k)$: each operation creates a constant number of new turning points.
	In particular, the total number of turning points is $\Oh(k)$, so manipulating BSTs takes $\Oh(\log k)$ time.
	Next, we note that the worst-case running time of most operations is $\Oh(\log k)$, with extra $\Oh(\log k)$
	time needed for each discarded or collapsed segment. However, every such segment decreases the potential by $\log k$. 
\end{proof}

\section{An $\Oh(mn\log(mn))$ time RLE Edit Distance Algorithm}
As in the previous algorithm by Chen and Chao~\cite{CC-RLEedit:2013}, we go through the dynamic programming table block by block.
For every block, we transform the representation of its \inborder to the representation of its \outborder. As mentioned earlier,
borders are piecewise linear with gradient $\pm 1$ or $0$ so they can be maintained in the structure described
in \cref{sec:newalgo}. We will assume that the left and the top part of the \inborder of every block are stored in separate structures.
We start by generating the structures corresponding to the left and the top border of the whole dynamic programming
table. 
These left and top borders are each a single decreasing and increasing sequence, respectively. As a result, we can generate the data structure for the parts corresponding to all blocks trivially  in $\Oh(m+n)$ time
using $m+n$ \textsf{create} operations.
%The corresponding sequences consist of a single linearly increasing segment and so this step is trivial to implement in $\Oh(m+n)$ time
%using $m+n$ operations \textsf{create}. 
Now, we have to describe how to obtain the structure corresponding to
the right and the bottom part of the \outborder of the current block given the structures corresponding to the left and the
top part of its \inborder. We stress that any structure will be created and then used once as an input to a further operation,
which is crucial for the amortisation argument within \cref{lem:overall}.

Recall that the semantics of \textsf{split} and \textsf{join} operating on arrays in \cref{sec:previous} and of \textsf{split} and \textsf{join} operating on piecewise
linear functions in \cref{sec:newalgo} is slightly different: \textsf{split} now creates two functions
that both contain the value of the original function at $x_{M}$; symmetrically, \textsf{join} takes two functions
defined on $[x_{L},x_{M}]$ and $[x_{M},x_{R}]$ that share the same value at $x_{M}$. This is, however, not an issue because
the cases in both \eqref{eqn:left} and \eqref{eqn:top} overlap at the boundaries.

For a match block, the value stored in an element $(i,j)$ of the \outborder is a copy of the value stored in the corresponding
element $(i_{d},j_{d})$ of the \inborder. Recalling that $(i_{d},j_{d})$ is the intersection of the \inborder with diagonal $j-i$,
this can be readily implemented with a constant number of \textsf{split} and \textsf{join} operations.

For a mismatch block, we need to apply \cref{alg:left,alg:top}, merge the returned solutions
by taking the minimum at every position, and finally create separate structures corresponding to the right and the bottom
part of the \outborder with a single \textsf{split} operation. Note that while we have already observed that both \inborder
and \outborder are piecewise linear with gradient $\pm 1$ or $0$, we need to make sure that the same is true for every
function obtained inside \cref{alg:left,alg:top}, and for \outputtop and \outputleft in particular.

\begin{lemma}
\label{lem:partial}
Every function obtained in \cref{alg:left,alg:top} is piecewise linear
with gradient $\pm 1$ or $0$.
\end{lemma}

\begin{proof}
Consider \cref{alg:left}. It is easy to verify that $S$ and hence also $S^{\ell}$ and $S^{r}$
are indeed piecewise linear with gradient $\pm 1$ or $0$. Additionally, $S^{\ell}[i]$ is equal
to the minimum in $\inputleft[1,i]$ and so $S^{\ell}$ is non-increasing. Consequently, $S_{1}$,
$S_{2}$, and $S_{3}$ are all piecewise linear with gradient $\pm 1$ or $0$. We only need to verify
that the same holds for their concatenation. This is true because each of these three parts
corresponds to a case considered in  \eqref{eqn:left}, and these cases overlap
at the boundaries.

Next, consider \cref{alg:top}. Similarly as above, it is easy to verify that $S$ and
so also $S^{\ell}$ and $S^{r}$ are piecewise linear with gradient $\pm 1$ or $0$. Furthermore,
$S^{r}[i]$ is equal to the minimum in $\inputtop[w-h+i+1,w]$ and so $S^{r}$ is non-decreasing.
Thus, $S_{1}$ and $S_{2}$ are piecewise linear with gradient $\pm 1$ or $0$ and the same holds
for their concatenation because the cases in \eqref{eqn:top} overlap at the boundaries.
\end{proof}

We now explain in detail how to implement \cref{alg:left}. We start with computing $S^{\ell}$ and $S^{r}$ by
first calling $\mathsf{SWM}(\inputleft,h-1)$ and then using \textsf{split}. Next, $S_{1}$ is obtained by applying
\textsf{gradient change} and \textsf{shift} to $S^{\ell}$, $S_{2}$ is obtained by calling \textsf{create}, and $S_{3}$ is obtained
by applying \textsf{shift} to $S^{r}$. Finally, \outputleft is created with two calls to \textsf{join}.

\cref{alg:top} is implemented by calling $\mathsf{SWM}(\inputtop,w-1)$ and then using \textsf{split}. Next, $S_{1}$
is obtained by applying \textsf{shift} to $S^{\ell}$, while $S_{2}$ is obtained by applying \textsf{gradient change} and \textsf{shift}
to $S^{r}$. Finally, \outputtop is created by a single call to \textsf{join}.

Having obtained a representation of \outputleft and \outputtop, we call \textsf{combine} to obtain a representation
of the \outborder. Such a call is valid due to \cref{lem:cross}.
The overall number of operations on structures is $\Oh(mn)$, making the final time complexity
$\Oh(m n\log(mn))$ by \cref{lem:overall}.

\bibliographystyle{plainurl}
\bibliography{rle}
\end{document}